# Exploring the use of time-varying graphs for modelling transit networks

Ikechukwu Maduako    Emerson Cavalheri    Monica Wachowicz
imaduako@unb.ca    e.cavalheri@unb.ca    monicaw@unb.ca

People in Motion Lab
University of New Brunswick
15 Dineen Drive, Fredericton, NB. E3B 5A3 Canada

**Abstract**

The study of the dynamic relationship between topological structure of a transit network and the mobility patterns of transit vehicles on this network is critical towards devising smart and time-aware solutions to transit management and recommendation systems. This paper proposes a time-varying graph (TVG) to model this relationship. The effectiveness of this proposed model has been explored by implementing the model in Neo4j graph database using transit feeds generated by bus transit network of the City of Moncton, New Brunswick, Canada. Dynamics in this relationship also have been detected using network metrics such as temporal shortest paths, degree, betweenness and PageRank centralities as well as temporal network diameter and density.

## 1. Introduction

Understanding the dynamic relationships between the topological structure of a transit network and the mobility patterns of transit vehicles on this network can help researchers to bring transit management a step forward into devising solutions to reliability, optimization, vulnerability and forecasting issues. The topological structure of the transit network represents the transit features such as streets, roads, highways, rail ways, metro-lines, waterways, stations and stops while the mobility characteristics are generated by the dynamic features on the network such as the buses, cars, trains, bikes and pedestrians. Exploring this relationships through graph analytics is very challenging and demands transportation researchers to have good capacity to handle gigantic amount of time-varying transit data (Lin & Ban, 2013). Not only that there is massive amount of data to be processed as a graph but modeling the graph to capture the dynamic topological and mobility pattern in time and space is another challenge. Many previous research has focused on using static dual and primal graphs to model the topological structure of transit network. On the other hand, some studies have represented mobility in transit network as trajectories of transit vehicles which are later matched with the edges of a static primal graph. Static graphs represent finite number of nodes and edges that are constant through time. Hence, they lack the means to capture the changes in the topological properties of the physical transit network structure as well as the dynamics of the mobility pattern (traffic flow, congestions) within the network. In view of this, time-varying graphs (TVG) have attracted attentions recently for modeling temporal changes in dynamic networks such as in social, biological, citation and web networks. A TVG represents an infinite number of nodes and edges that constantly change in terms of structure and connectivity through time. Time-varying graphs have been adopted in different graph problems where temporal context and dynamic relationships in the graph are of interest such as clustering, classification, influence analysis, link prediction, tensor analysis, community building, outlier detection, event detection, and evolution of graph metrics of real-world networks. However, To the best of our knowledge, there is no study applying time varying graphs to capture the dynamics in the relationship between the topological structure and the mobility patterns of a transit network Therefore, this paper proposes a time-varying graph model (TVG) for modelling and analyzing the dynamic relationship between topological structures and mobility patterns in a transit network. This TVG model provides us with the capability to carry out topological structural computations (e.g. shortest paths, network diameter) as well as mobility pattern detection (e.g. traffic flow, congestions) in a transit network

This paper proceeds as follows; Section 2 provides a review of related work on modeling and analysis of transit network as graphs as well as studies utilizing TVG to model relationships and dynamics in real world networks. Section 3 describes our TVG model for transit network. Implemented graph metrics for the graph analytics are defined in section 4. The practical implementation of the model in Neo4j graph database is described in section 5 while analysis and results are provided in section 6. Finally, section 7 concludes the paper.

## 2. Related Work

Late 1980s and 1990s emerged the use of graphs concepts in the study and analysis of public transit networks (Derrible & Kennedy, 2011). Till date, studies on transit networks utilizing graphs have focused on the physical structure of the network based on static graph models for purposes of design and study of the structure of the network. Traditional approaches such as primal and dual graph representations have been mostly used in the past to model transit networks (Aǹez et al.1996). This section reviews some of these previous works on transit network analysis with graphs as well as studies that utilized TVG model in other real world networks.

Anez et al. (1994) focused on modeling transit route and understanding the complexity level introduced by dual graph representation in transit networks. The conclusion is that modeling intersections as links and streets as nodes creates fewer nodes and computationally more efficient than traditional primal model. However, this was based on static dual graph representation, hence dynamic relationships in the graph entities of the transit network used in the work cannot be analyzed with the model for lack of temporal context in the model. Other examples include, Porta et al. (2009) who studied the relationship between street centrality and densities of commercial activities in the city of Bologna, Italy utilizing a static primal graph model. This work focused on static physical structure of street networks and its influence on commercial activities in Bologna. Wang et al. (2011) analyzed the relationship between transit network structure and nodal centrality of individual cities in China based on static primal graph model. Similar study was carried out in (F. Wang, Antipova, & Porta, 2011) to examine the relationship between street centrality and land use intensity in Balton Rouge, Louisiana. Graph metrics used in this study were closeness, betweenness and straightness centralities of the street (road) network based on static primal graph model. Derrible (2012) carried out betweenness centrality analysis on 28 metro systems worldwide based on static primal graph model. The study focused more on the topological structure of the network and on how this property varies with network size comparing different metro systems worldwide. Static graph analysis was carried to measure performance and estimate resilience of Australian air transit network by (Hossain, Alam, Rees, & Abbass, 2013). In this study, real air traffic data for all domestic flight in 2011 was used to build a primal graph model of the air transit network. Graph metric analysis was carried out looking for the connectedness of the network, flight path and reachability of the airport nodes as well as degree and betweenness centralities of the airport nodes.

In summary, the comprehensive survey of (Derrible & Kennedy, 2011) as well as that of Lin and Ban (2013) on graph theory and applications to transit network also agreed that previous works have focused on static graph measures of topological structure of transit network. No attempts have been made to explore the dynamic relationships between underlying transit network structure and mobility pattern on it based on graph modeling and analysis. However, the traditional graph models and approaches reviewed in this section cannot be used to model and analyze these dynamic relationships mainly because, (i), the models are based on static graphs whereas mobility pattern is dynamic, (ii), the models focused mainly on the physical transit network characteristics, (iii), the models are not suitable for multi-modal integration and (iv), while the approaches that are based on trajectories of transit vehicles matched on primal graph are not efficient for computation of most important graph metrics to analyze the relationships in transit network including distance or time-weighted graph metrics such as degree and PageRank centralities.

Therefore, Time-varying graph (TVG) presents the means to model dynamic relationships in transit networks and track the changes overtime as has been effectively utilized in other complex networks. Studies utilizing time-varying graphs have concentrated on social, web, citation and biological networks and two major approaches of TVG have been mostly used. These are Sequence of Snapshots method and Aggregated Graph method.

On one hand, snapshots method models dynamic networks as a sequence of graph snapshots, $G_{[t1-tn]} = \{$

$G_1, G_2, G_3, \ldots, G_n\}$. Each graph snapshot $G_i$ is a static graph that represents a valid state of a network at time point $t_i$ (Zaki & Attia, 2016). This methods have been adopted in the works of (Köhler, Langkau, & Skutella, 2002), (Tantipathananandh, Berger-Wolf, & Kempe, 2007), (Lahiri & Berger-Wolf, 2008), (Rossi, Gallagher, Neville, & Henderson, 2013), (Yang, Yu, Gao, Pei, & Li, 2014), (Khurana & Deshpande, 2016) and (Semertzidis & Pitoura, 2016). However, several limitations arise when using this approach in fast-streaming data such as in transit networks. These limitations range from high storage overhead due to the replication of an entire graph every time there is an update or change, to the expensive computation of longitudinal queries across the snapshots. On the other hand, Aggregated graph method models a dynamic network as one large time-indexed graph, $G_{[t1,tn]} = [N_{[t1-tn]}, E_{[t1-tn]}]$, where $N_{[t1-tn]}$ and $E_{[t1-tn]}$ are sets of all node and edge instances respectively (Zaki & Attia, 2016). Betsy and Shashi (2009), Papastefanatos, Stavrakas and Galani (2013), Cattuto et al. (2013) and Huo and Tsotras (2014) have adopted this approach to model time-varying social network graphs. We have adapted this approach to model transit network data as a large aggregated graph indexed on a hierarchical time-tree temporal indexing structure. The main reason is to reduce replication of nodes and ability to carry out longitudinal queries across time more efficiently as opposed to sequence of snapshot approach.

## 3. Main components of our Transit Time-Varying Graph

The proposed Time-Varying Graph model for transit network is a directed graph model with two key aspects. The vertices (entities of the network) and the time instants (time-tree) are the key features (aspects) of the graph.

The graph model is defined as G = (N, E, T): where N is the set of nodes (network entities), E is the set of all edges (network relationships) and T is the set of all time instants (time-tree).

The set of all nodes in G is denoted as N(G), the set of all edges as E(G) and the set of all time instants as T(G). N(G) is made up of spatial node $n_s$, temporal node $n_t$ and spatio-temporal node $n_{st}$ instants. E(G) is consists of spatial edge $e_s$, temporal edge $e_t$ and spatio-temporal edge $e_{st}$ instants. A spatial node in the graph model is a node that have spatial coordinates among other properties that it has but there is no temporal property, for example bus stop node, route node and street node. Temporal nodes are nodes with timestamps and have leaf nodes connected to the time-tree but have no spatial properties, for example the trip node. While the spatio-temporal nodes are nodes with spatial and temporal properties in the transit graph model, for example, Move and Stop nodes. Spatial edges have spatial edge weights (distance) but no temporal property, an example is the edge between bus stop node and street segment node. Temporal edges are edges between two temporal nodes but not spatially weighted, an example of this, is the edge between Trip and TripOrigin node as well as arrival and departure edges to the bus stops, they are weighted by time. Spatio-temporal edges are directed edges between two spatio-temporal nodes, it has temporal attributes as well as spatially weighted, an example is the edge between Moves and Stops nodes (the NEXT edges).

A spatial edge $e_s \in E(G)$ is defined as an ordered triple $e_s = (u, v, w_s)$, where u, v $\in$ N(G) are the source and target nodes, respectively and $w_s$ is the spatial weight of the edge (distance).

A temporal edge $e_t \in E(G)$ is defined as an ordered quintuple, $e_t = (u, t_a, v, t_b, w_t)$, where u, v $\in$ N(G) are the source and target nodes, respectively, $t_a, t_b \in$ T(G) are the source and target time instants, respectively and $w_t$ is the temporal weight of the edge (time).

A spatio-temporal edge $e_{st} \in E(G)$ is defined as an ordered sextuple $e_{st} = (u, t_a, v, t_b, w_s, w_t)$ where u, v $\in$ N(G) are the source and target nodes, respectively, $t_a, t_b \in$ T(G) are the source and target time instants, respectively, $w_s$ is the spatial weight of the edge (distance) and $w_t$ is the temporal weight of the edge (time)

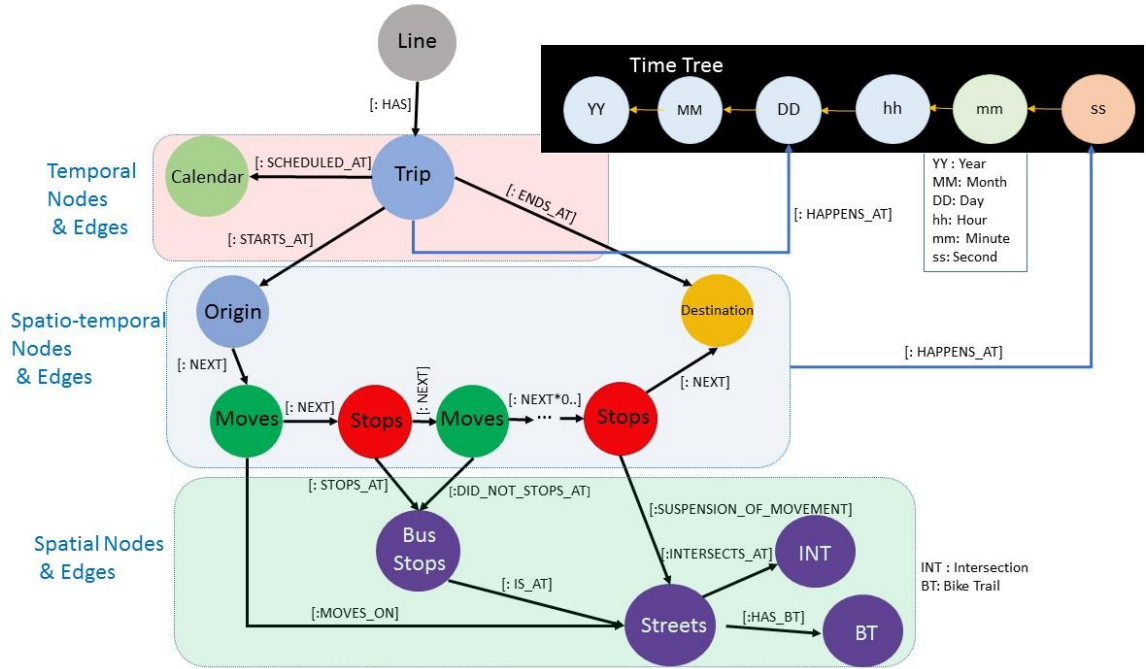

*Figure 1. Logical Graph Model of the TVG for Transit Network.*

Figure 1 shows the logical graph model of the transit network TVG. In this model, the graph objects (node and relationships) are represented with their respective semantic labels and grouping. the Transit network entities are represented as nodes and their relationships as links in contrast to what is obtained from dual and primal representations. The nodes are spatially and temporally indexed to enable distance-weighted metrics computation and time-dependent analysis on the graph respectively. Temporal and spatio-temporal nodes are connected to the time tree which also keeps track of the changes in the graph. The finest unit of temporal aggregation is the 'second' time granularity and spatial is the vector point.

A Line (route) HAS many trips per day. A Trip is SCHEDULED_AT according to the transit Calendar and STARTS_AT an Origin (bus stop or street) and ENDS_AT a Destination (bus stop or street). A trip from an Origin node to a Destination node, having a sequence of Moves and Stops nodes linked by the NEXT relationship. The Moves are spatially connected to the Street Segment on which a bus is moving. Stops can be connected to Bus Stops at which a bus stops and other Stops can be connected to the Street Segment or Street Intersection. There are some Moves that are to the Bus Stops, indicating the cases where they should have stopped at the Bus Stops but they did not stop, the relationship is semantically labeled in the model as DID_NOT_STOP_AT. The edges in the graph model are directed, the NEXT edges between Move and Stop or Move and Move are weighted by the distances between them, enabling Dijkstra shortest path computation.

Nodes and edges with temporal properties are connected to the leaf nodes of the time-tree instants. With this approach, time-varying graph queries or simply finding events that occurred at a specific time or within a range of time are enabled. Any time-dependent query firstly arrives at the time-instant node or range of time-instant nodes of the time-tree and traverses to all events or nodes that are linked to it without having to scan through all the nodes of the large graph.

The time-tree has a Root node that contains the Year level nodes, the Year node contains the Month level nodes and so on, it can be as granular as needed. The top or lower level nodes can be reached through the traversal of the CONTAINS edges. Nodes at each level of the time-tree are sequentially connected through the NEXT relationship as depicted in figure 2.

### 4. Network Metrics

Dynamic network metrics such as connectedness, time-dependent shortest path, degree, betweenness and PageRank centralities as well as network diameter and density and their practical implications to transit management are the major focus and the reasons are explained in the definitions of the metrics provided in the following section. These metrics are implemented and retrieved through time-dependent graph database ad-hoc queries in Neo4j platform and the results are visualized as graphs, charts and tables. These ad-hoc queries are implemented over time-points (such as peak hours) or time-intervals (such as morning, afternoon, evening or daily interval) using the time-tree.

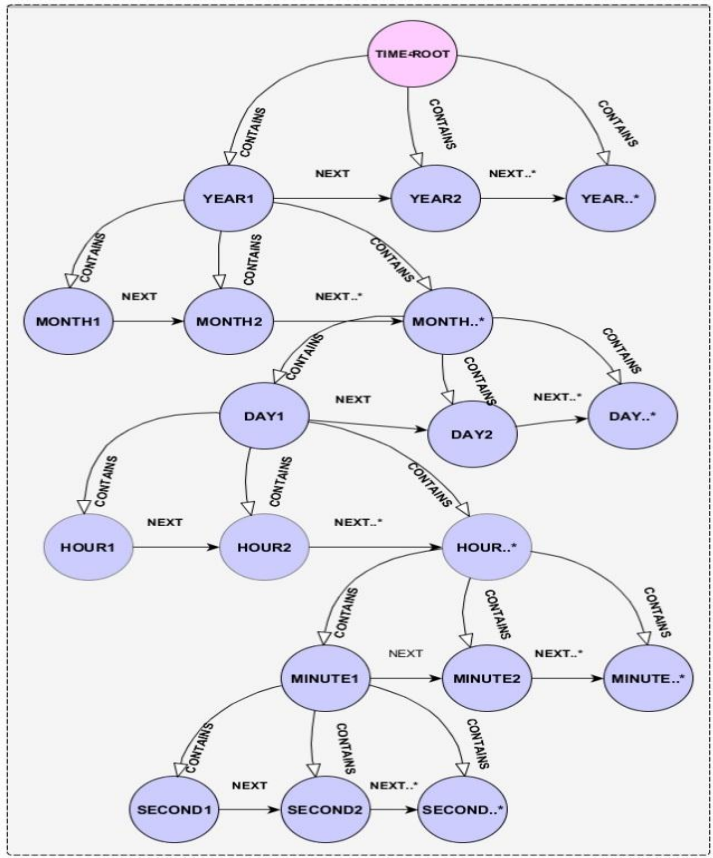

*Figure 2: Time-Tree indexing structure*

## 4.1. Definitions

### 4.1.1. Connectivity (connectedness)

Connectivity is one of the key benefits of graph database over relational database. A graph is said to be connected if there is a path between every pair of vertex. There is always a path to traverse from one vertex to another. Graph is said to be disconnected when there is a or multiple vertices and edges that are disjointed. In the context of time-varying transit network graph, the connectivity of bus trips from origins to destinations over time is of major interest. Structural behaviors and anomalies in the bus trips over time is the major analysis.

### 4.1.2. Path (shortest and longest shortest paths)

This is one of the most common metric in graph. A path in graph is simply a sequence of nodes with the property that each consecutive pair in the sequence is connected by weighted or unweighted edge. Things travel or traverse across the edges of the graph from an origin node to a destination node. This could be a bus trip across the edges connecting an origin to a destination with sequence of moves and stops or pedestrian taking a sequence of moves and stops from an origin to a destination. In this paper, time-varying shortest trips and longest shortest trips in the transit network leveraging shortest and longest shortest path algorithms is of interest.

### 4.1.3. Centralities (degree, betweenness and PageRank)

Centrality measures give relative measure of importance in the network. The different centrality measures have different type of metric of importance.

### 4.1.4. Degree centrality

Degree of a node is simply the number of connections (the in-degree and out-degrees) that the node has in the network. The higher the number of connections the more influential the node has in the network. In this paper, degree centrality algorithm is leveraged to retrieve the busiest streets, lines and stops over time in the transit network.

### 4.1.5. Betweenness centrality

Betweenness centrality of a node in the network is the number of shortest paths between two other nodes in the network on which a given node appears. This is an important metric in the network because it can be used to identify nodes that can server as "broker of services" in the network. In the context of time-varying transit network analysis, time-dependent transfer points are identified utilizing betweenness centrality measure over time points and time intervals.

### 4.1.6. PageRank centrality

PageRank was originally used by Google (Brin & Page, 1998) to rank importance of search results in the web network. This centrality measure is similar to eigenvector centrality which scores nodes in the network according to the number of connections to high-degree nodes they have. The results from PageRank centrality analysis most times correlates with that of degree centrality as both measure the level of importance of a node based on connectedness. However, PageRank considers nodes as more important, if they have more incoming edges than outgoing ones as well as the importance of the nodes that link to it. In this paper, PageRank is leveraged to analysis level of importance, the streets, lines and stops in the transit network have at different time points and intervals.

### 4.1.7. Network Diameter and Density

Diameter (geodesic) of a network is the longest shortest path in the network. In this paper, we leveraged the longest shortest path algorithm to compute the diameter of the transit network at different time points and intervals to identify longest trips in the network across different times.

Network density is a measure how close the graph is to a complete graph. The density of a graph G = (N, E) is defined as: $D = \frac{|E|}{|N|*(|N|-1)}$ where N is the total number of nodes and E is the total number of edges (links). In this paper, density of the transit network is computed over different time points and time intervals to identify peak and less peak periods in the network.

## 5. Implementation

The TVG was implemented in Neo4j graph database management system. Neo4j is currently the most popular native graph database widely used with a relatively stable open source version. Database development and query language in Neo4j is called Cypher which is as declarative as SQL for relational database. Cypher enables vast analytical graph algorithms and ad-hoc queries as well as user defined functions (UDFs).

The time-varying graph database starts with the building of the time-tree from year to second levels of aggregation. Secondly the graph entities (nodes, attributes and relationships) were coded and loaded into the graph database engine. The temporal nodes of the graph are sequentially connected to the time-tree from top to the lowest level of time granularity, an example is shown in listing 1 which describes the sequential connection of the Move nodes to the time-tree leaf nodes from time granularity of year to seconds. The sequential NEXT relationship edges between consecutive Move and Move or Move and Stop are weighted by their respective distances to enable further analysis.

Listing 1: Cypher coded- Sequential connection of Move nodes to the Time-Tree

*MATCH (t:Moves) WITH t*
*MATCH (yy:Year {yearid:t.year}) WITH t,yy*
*MATCH (yy)-[r1]->(mm:Month {monthid:t.month}) WITH t,yy,mm*
*MATCH (mm)-[r2]->(dd:Day {dayid:t.day}) WITH t,yy,mm,dd*
*MATCH (dd)-[r3]->(hh:Hour {hourid:t.hour}) WITH t,yy,mm,dd, hh*
*MATCH (hh)-[r4]->(mm1:Minute {minuteid:t.minute}) WITH t,yy,mm,dd, hh, mm1*
*MATCH (mm1)-[r5]->(ss:Second {secondid:t.second}) WITH t,yy,mm,dd,hh,mm1,ss*
*CREATE (t)-[:HAPPENS_AT]->(ss);*

### 5.1. Dataset

The General Transit Feed Specification (GTFS) data has been widely used in transit networks analysis and time -dependent accessibility studies such as in (Farber, Morang, & Widener, 2014), (Fransen et al., 2015), (Farber, Ritter, & Fu, 2016)  as wells as in real-time movement visualization, (Bast, Brosi, & Storandt, 2014). However, GTFS data is not reliable for the uncertainties in  travel, departure and arrival time as well as other uncertainties caused by operational delay, service interruptions and unrealistic schedules (Wessel, Allen, & Farber, 2016). Therefore, real-time accessibility of GPS location feeds from the transit system is more desirable and accurate. We have used a two-week transit data feeds of the Codiac Transpo Moncton buses from their location service web API.  (explain more here, the GPS coordinates were retrieved every 5 seconds, etc.) In addition, freely available transit network spatial datasets of bus route, bus stop locations, street segments, street intersections locations and civic addresses were downloaded from GeoNB online service (http://www.snb.ca/geonb1/e/index-E.asp) for data contextualization. The datasets went through a pipeline of data pre-processing, transformation and aggregation to make them ready to be used in building the transit network graph (Cao & Wachowicz, 2017).

## 6. Analysis Results

The graph metric algorithms defined in the section 4.1 are implemented and called inside cypher ad-hoc queries to retrieve results from the transit network graph database in Neo4j. Longitudinal queries across time used for temporal analysis are implemented as time-incremental cypher queries referring to the time-tree. An example of such queries is in listing 2 below.

Listing 2: Cypher coded- metric algorithm in time-incremental query example.

*WITH range (6,23) AS Hour*
*FOREACH (hour IN Hour |*
*CALL PageRank (({*
 *Write:True,*
 *Property: 'pagerank',*
*Node:'none',*
*Relationship:'MATCH (d: Day)- [: HAS_HOUR]-(h:Hour)<-[r1:HAPPENS_AT]-(st:Moves)*
*WITH st*
*MATCH (st)-[r]-(bs:Streets)*
*RETURN id(st) AS source, id(bs) AS target, count (\*) AS weight ORDER BY weight DESC LIMIT 20'}))*

6.1. Trip connectivity Analysis

Figure 3 is a graph visualization of the pattern of connectedness made by a bus trip (50-12) with other graph entities in the network from its origin to its destination over a bus route at 8 am in the morning across 4 consecutive days. This pattern is constantly changing based on the overall traffic situation on the bus route. The graph visualization makes it easier for transit managers to visualize the stops that were made at the bus stops or at the intersections at different point in time. Where and when the trip experienced some stops due to traffic congestion, they are also easily spotted out from the Connectivity is one of the key benefits of graph database over relational database.

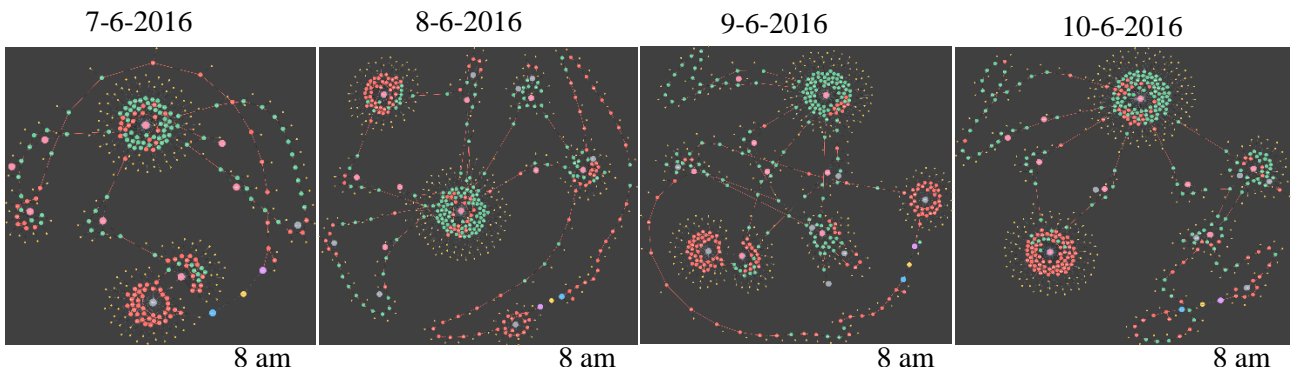

Figure 3: Mobility patterns of trip 50-12 across 4 consecutive days. Red nodes: Stops, Green nodes: Moves, Pink nodes: Streets: Gray nodes: Bus stops, Blue and Purple nodes are Origin and Destination points of the trip and the tiny orange nodes are time instants from the time tree.

Transit managers could visualize the dynamic behavior of this trip over time in addition to the attribute information of these entities. Anomalies in the network of trips can be also monitored and detected from ad-hoc queries to the transit network graph database. Integrating structural information and observational values of the graph entities exposes more insight about the dynamics of the network.

6.2. Path metrics

Path queries and analysis are very useful information for fleet management based on shortest or longest shortest path algorithm in the graph database. From our developed time-varying transit network graph database, it is very insightful to carry out time-dependent shortest or longest shortest path queries across time points to detect for example longest trips and shortest trips at different times of the day. Figure 4 shows the graph of the longest and shortest bus trips at 8 am, 4 pm and 6 pm peak hours on the 9[th] of June 2016. The computation is based on Dijkstra shortest and longest shortest path algorithm with edges weighted by distance and time. The number of hops, distances and time covered by the trips are also provided. For the fleet managers, this provides informative insights on why some trips took longer and some spent shorter time at different hours of the day based on the composition of the graph.

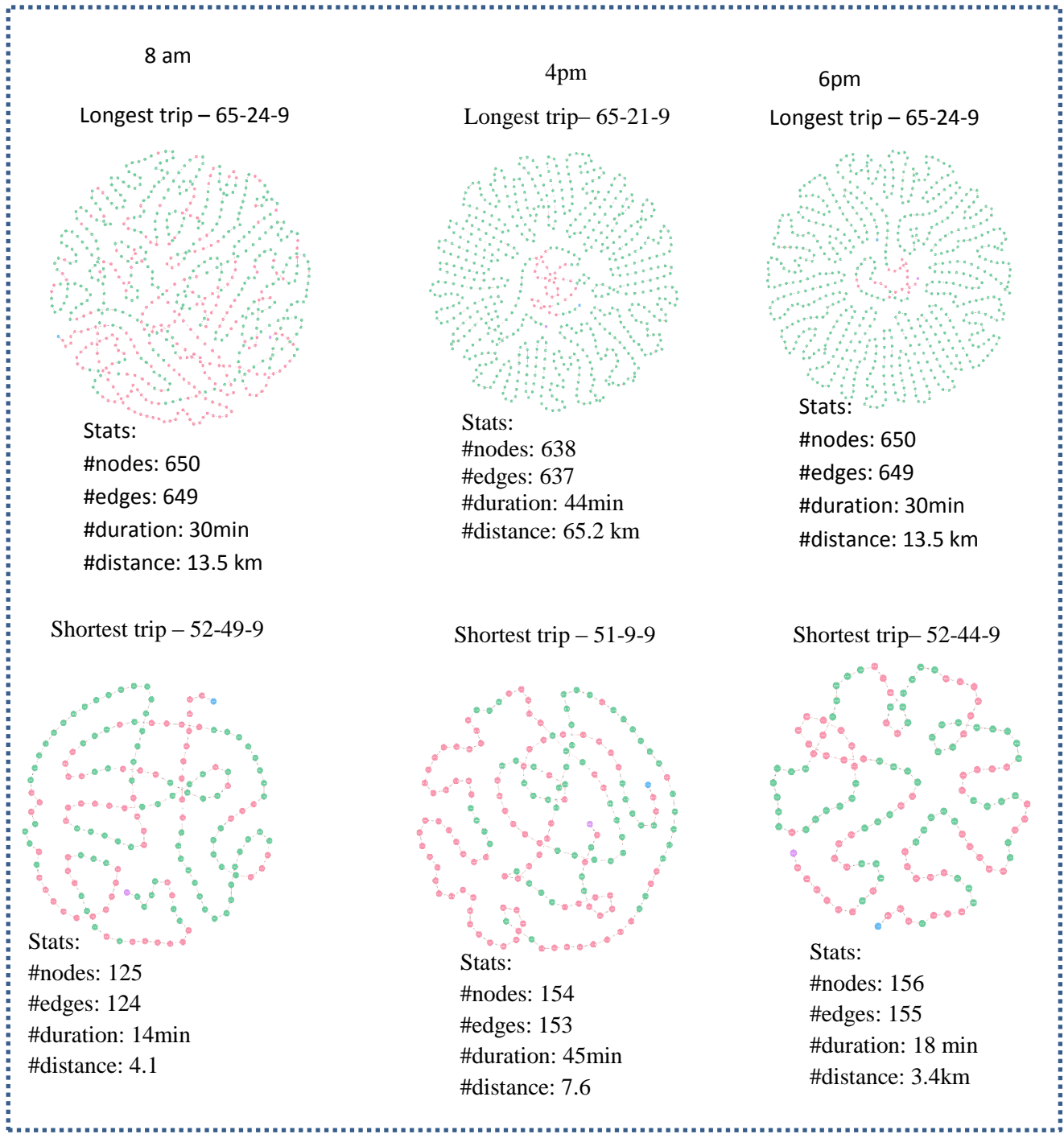

Figure 4: Time-varying Shortest and Longest trips at peak hours

### 6.3. Network Diameter

Figure 5 depicts the average network diameters of the transit network at peak hours of the day based on the longest shortest path algorithm. The diameter of the network is constantly changing across time due to the changes in mobility pattern in the network. The diameter in this context signifies the average longest trips at these hours. This is crucial for public transit managers especially for fleet planning and management.

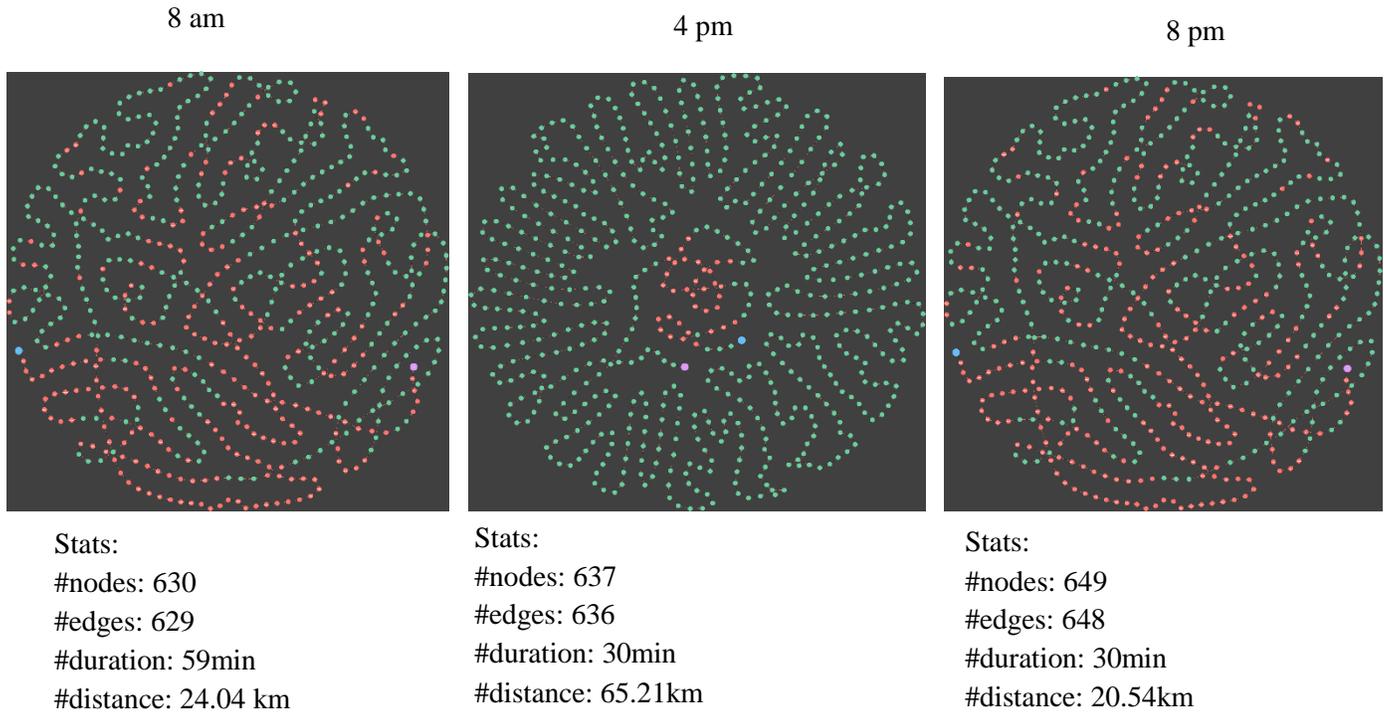

8 am
Stats:
#nodes: 630
#edges: 629
#duration: 59min
#distance: 24.04 km

4 pm
Stats:
#nodes: 637
#edges: 636
#duration: 30min
#distance: 65.21km

8 pm
Stats:
#nodes: 649
#edges: 648
#duration: 30min
#distance: 20.54km

Figure 5: Transit network average Diameter in terms of longest trips at peak hours

### 6.4. Centrality Metrics

Centrality measures give us relative measures of importance of any specific node in the network. In the context of our time-varying transit network graph, degree of centrality of the streets, bus stops and bus lines are measured, retrieved and visualized in forms of tables, charts and graphs, depicting their degrees of importance in the network. Figure 7 shows top 10 streets, bus stops and bus lines in Moncton with the highest degree of importance in terms of movements, stops and trips respectively. This correlates literally to the busiest streets, bus stops and bus lines in the network. Figure 8 shows the charts of the top busiest streets and bus stops in term of their degree of interactions in the network. The charts at the top depict the average degree over a period of 2 weeks and at the bottom are the charts depicting degrees across the hours of the day, from 6 am to 10 pm.  Main street has the highest number of movements and interaction in the network while the bus stop at Plaza Blvd (Walmart) records the highest number of stops per day and the line 51 in figure 7 records the highest number of trips.  The degrees of average level of suspension of movement (congestion) that could be experienced every day and every hour on those streets are shown in figure 6.

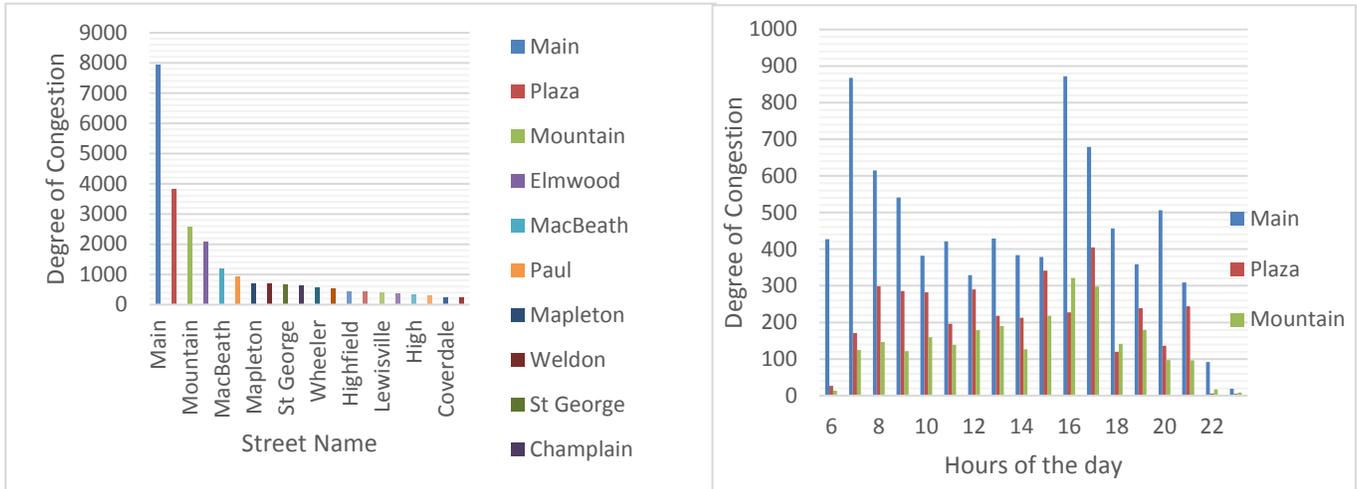

Figure 6: Average Degree of Congestion on the streets (left) and Varying-Degree of Congestion of the top 3 ranking streets across the hours of the day (right)

The chart on the left is the average degree of congestion over a period of 2 weeks of data collection. On the right is the average degree of congestion experienced every hour of the day from 6 am to 10 pm. This results were computed from degree of centrality algorithm over the connectivity of the street nodes and stop nodes based on. The hourly charts of the busiest streets and bus stops show some degree of fluctuations at different hours. Analysis over a longer period of time would certainly provide useful information for time-dependent transit route recommendation systems.

The degree of centrality ranking of the bus lines based on their number of trips is a very useful insight toward the decisions on which bus lines to merge or discard to reduce empty trips and which of the lines to optimize to improve ridership.

## Streets

| Street Name | Degree |
|---|---|
| Mountain | 10856 |
| Main | 8361 |
| Wheeler | 4027 |
| St George | 2654 |
| Champlain | 2507 |
| Plaza | 2302 |
| Lewisville | 2199 |
| Weldon | 2168 |
| Shediac | 2003 |
| Paul | 1683 |

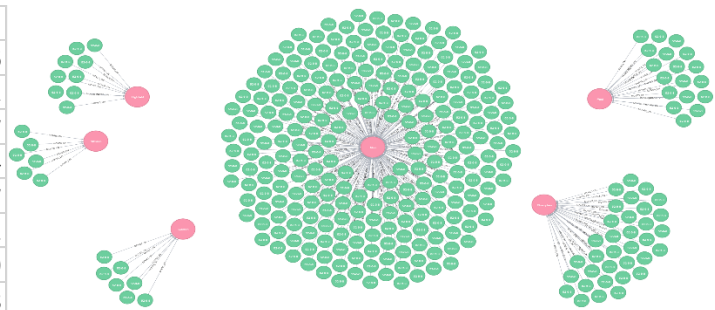

## Bus Stops

| Bus Stop Name | Degree |
|---|---|
| Plaza Blvd (Walmart) | 6533 |
| Champlain Place | 5527 |
| 1111 Main | 4851 |
| 54 Highfield | 532 |
| 77 Weldon | 468 |
| Riverview Place | 444 |
| 357 St George at Wel | 346 |
| 353 St George at Wel | 346 |
| 1045 Main (Highfield | 333 |
| 269 Weldon | 317 |

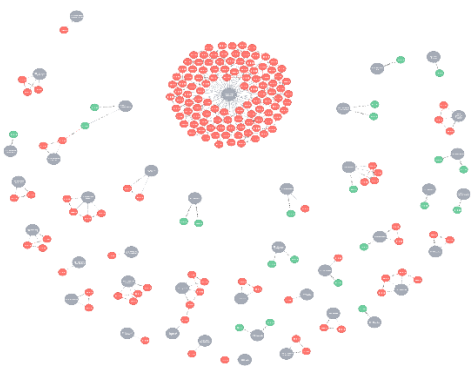

## Bus Lines

| Line Name | Degree |
|---|---|
| 51 | 206 |
| 52 | 186 |
| 94 | 116 |
| 63 | 101 |
| 64 | 101 |
| 50 | 99 |
| 65 | 98 |
| 61 | 98 |
| 62 | 97 |
| 60 | 94 |

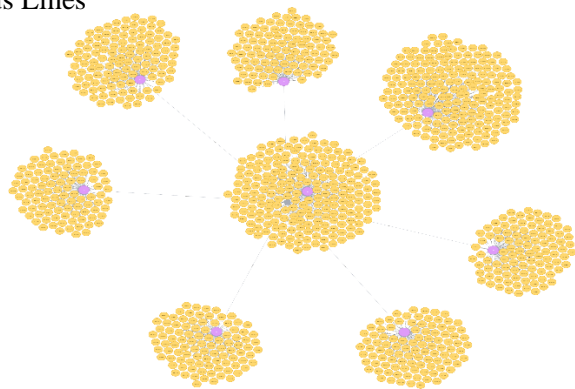

Figure 7: Table and Graph visualization of Degree Centralities of Streets, Bus Stops and Bus Lines respectively.

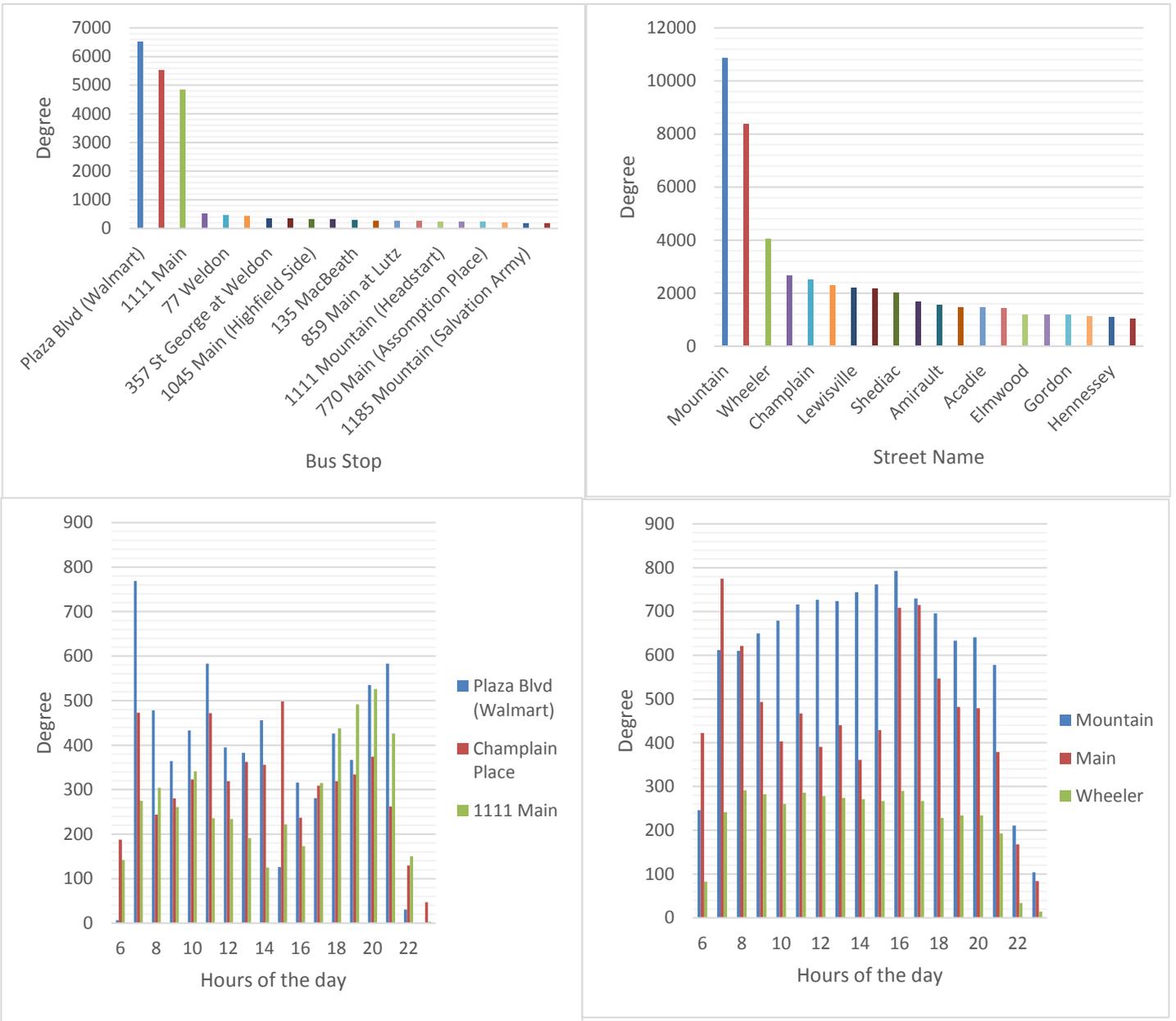

Figure 8: Average Degree Centrality charts of the bus stops and streets (top) and Time-Varying Centralities of the top 3 ranking bus stops and streets over the hours of the day (bottom)

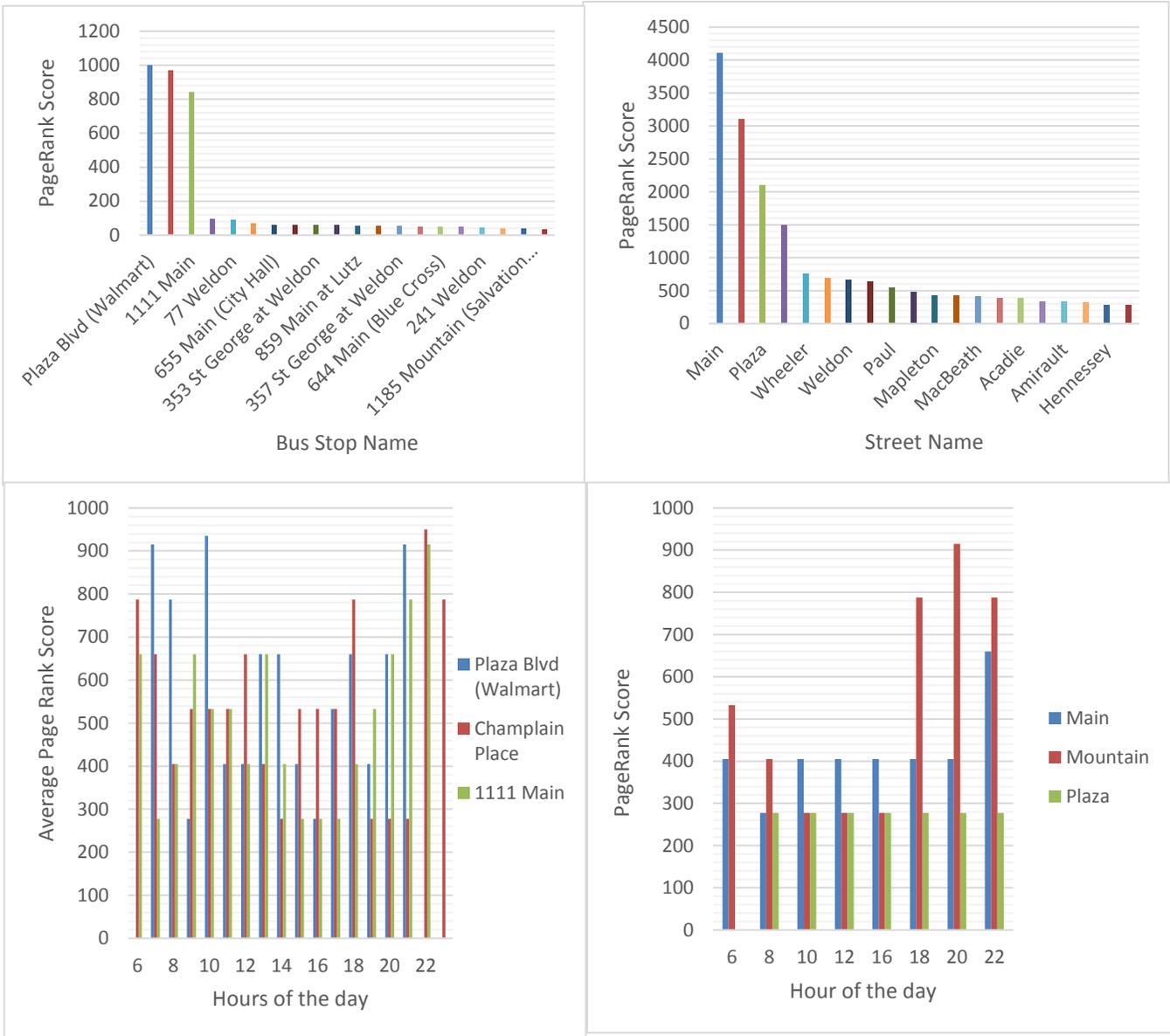

Figure 9: Average PageRank Centrality charts of the bus stops and streets (top) and Time-Varying PageRank of the top 3 ranking bus stops and streets respectively over the hours of the day (bottom)

The Page Rank analysis histograms that are shown in figure 9 provide the measure of popularity or importance similar to what degree of centrality has shown of the top 20 streets and bus stops in the transit network of Moncton. However, as degree of centrality utilizes both in-edges and out-edges for its computation, PageRank gives more importance to the in-coming edges and the importance of the nodes that link to the node in question. Main, Mountain and Plaza streets are the top 3 ranked streets in Moncton based on the amount of mobility on them. As well, bus stops at Walmart Plaza Blvd, Champlain and 1111 Main record the highest number of stops per day as also described by the degree of centrality charts. Similar changes we see on the hourly degree of centrality charts are shown by the hourly ranking of the streets and bus stops using PageRank algorithm. These are important decision making information for the transit planners and managers to prioritize and channel resources appropriately.

The extent to which a bus stop lies between other nodes in the network has been described in the betweenness centrality results in figures 10 and 11. A bus stop has higher betweenness score, if it is on the shortest paths connecting many other bus stops, hence it will be in the position to broker services or serve as a transfer point in the transit network. This transfer characteristic is a very important feature in transit networks. In this context, the importance of a bus stop does not only rely on its location e.g. within a business center or a mall (Plaza blvd bus stop) but even more important because it serves as a transfer point to get to many locations (Derrible, 2012). Figure 10 describes the average betweenness centrality of the top 20 busiest bus stops in the transit network. The Plaza Blvd (Walmart) bus stop is not only that it is located close to the mall but it also has the highest betweenness score because many other places can be reached through it. It is a very major transfer point in the transit network of Moncton.

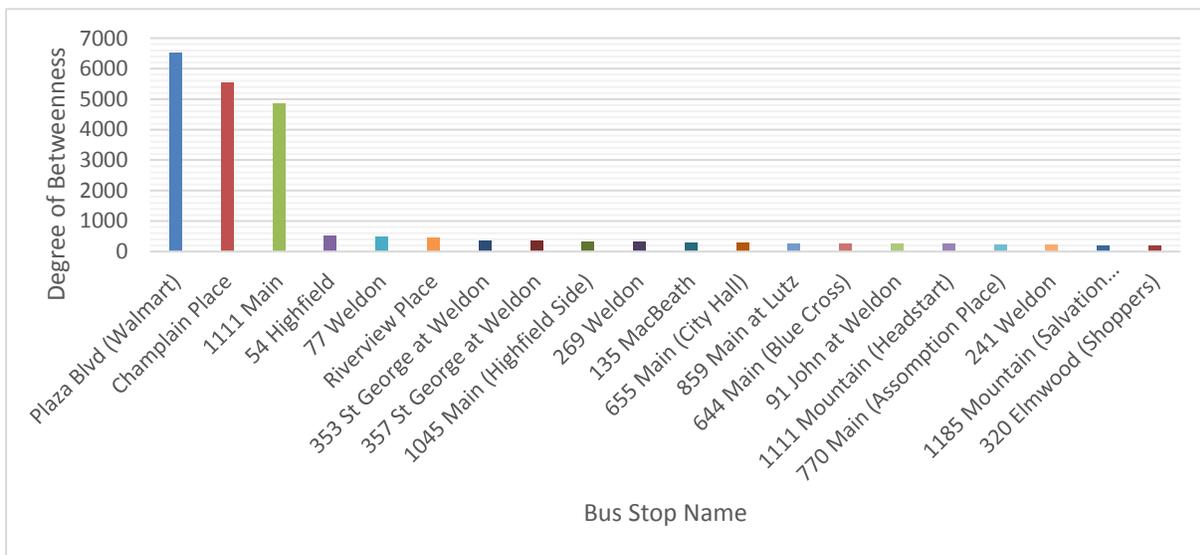

Figure 10: Average Betweenness Centrality charts of the top 20 busiest bus stops

However, based on dynamic mobility pattern in the network, this betweenness centrality of the bus stops changes from time to time. For example, Plaza Blvd (Walmart) bus stop is the most central at 8 am but not the most central at 8 pm as described by the charts in figure11. Time-dependent betweenness centrality computation is very important for smart transit system that is time-aware in the evenly distribution of passengers in the network

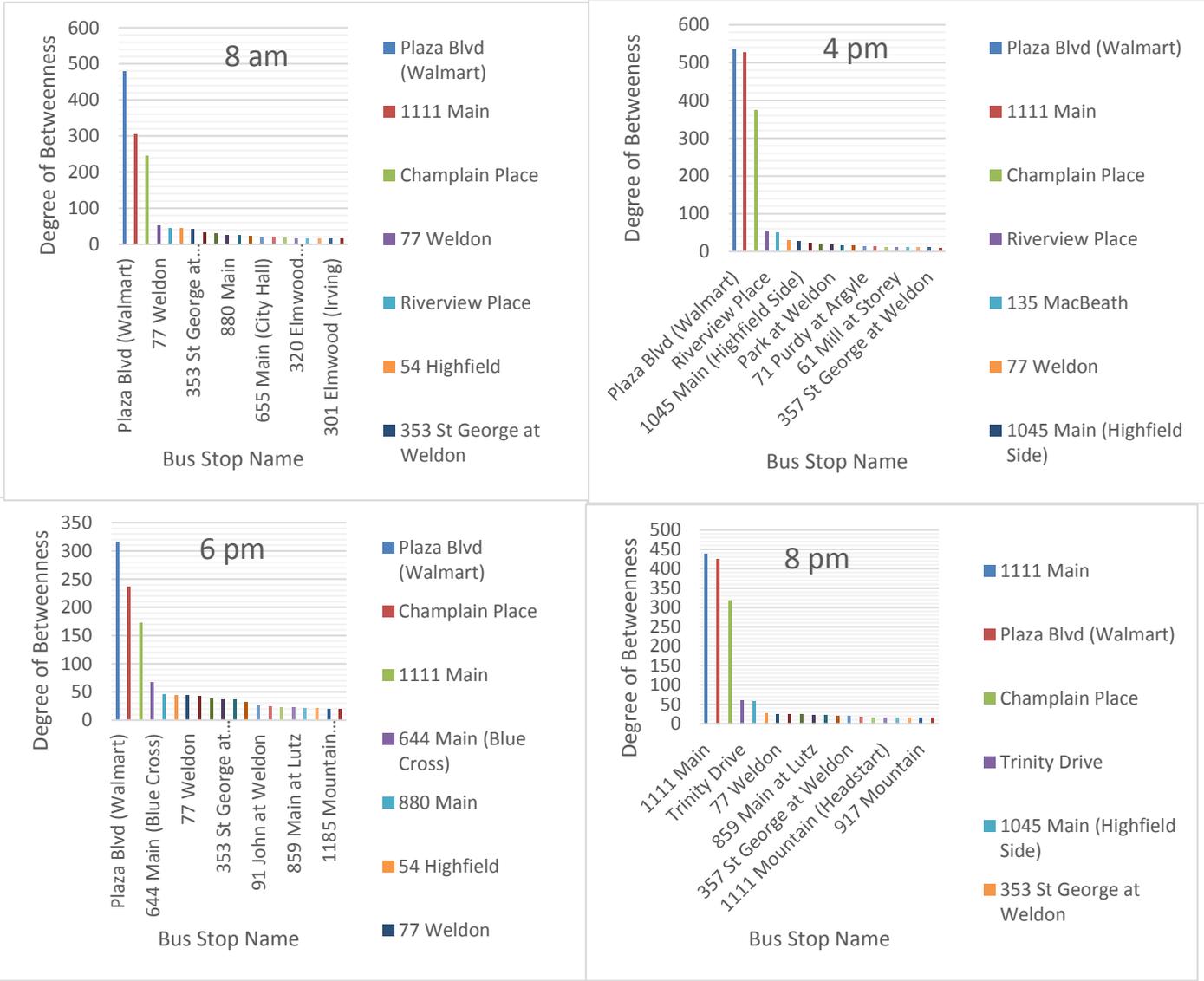

Figure 11: Time-Varying Betweenness Centrality of the Bus Stops at peak hours

## 7. Conclusion

While previous research on transit network analysis utilizing graphs have focused on traditional static graph models and technics with more emphasis on topological structure of transit network. In this paper, we have focused on understanding the dynamic relationships between topological structure of the transit network and mobility characteristics of public bus transit system on the network.

As mentioned earlier, understanding the dynamic relationships between the topological structure of a transit network and the mobility patterns of transit vehicles on it, is very vital to smart and time-aware transit design, management and recommendation systems.

This paper proposes time-varying graph (TVG) for modelling this dynamic relationship. The TVG model provides us with the capability to carry out topological structural computations (e.g. shortest paths, network diameter) as well as mobility pattern detection (e.g. traffic flow, congestions) in a transit network. This was implemented using a time tree in a native graph database, Neo4j. The time-tree is a hierarchical temporal index structure built on-demand that supports time indexing from year to milliseconds. In the TVG model, transit network entities are represented as nodes and the relationships between the entities are the links. Nodes include stops and moves of a moving transit vehicle, as well as street segments, intersections, trips, bus lines and bus stations of a transit network. Nodes with temporal properties are attached to the time instants of the time tree and are created when transit data feeds are inserted into the Neo4j database. Links are used to represent dynamic relationships such as next, now, before, and after.

We explored the effectiveness of the TVG model by implementing it using the transit feeds generated by the bus transit network of the City of Moncton, New Brunswick, Canada. Mobility patterns have been detected using network metrics such as temporal shortest paths, degree, betweenness and PageRank centralities as well as temporal network diameter and density. Shortest trips and longest trips found in the data provide transit managers with new insights on why some trips took longer or shorter time than usual at particular street segments and intersections. Furthermore, time-dependent degree and PageRank centrality have shown the busiest streets, bus stations and bus routes at different times of the day and days of the week. Temporal shortest paths and trip connectivity patterns were also detected together with betweenness centralities of the transfer points at peak hours of the day, leading to new insights on the location of traffic congestions on different streets and bus stations over time.

The next phase of this work includes scaling up the graph database with more than one year data feeds to assess the performance of the graph model in dealing with massive data and large graph processing. In addition, a detailed analysis of each of the transit graph metrics over a longer period of time will be carried out to enable predictive and evolutionary insights.